\documentclass{paper}
\usepackage{amsmath, graphics,
mathrsfs,latexsym,amscd,graphicx,amssymb, bm, upgreek}
\newcommand{\be}{\begin{equation}}
\newcommand{\ee}{\end{equation}}
\newcommand{\ba}{\begin{eqnarray}}
\newcommand{\ea}{\end{eqnarray}}

\newcommand{\pa}{\partial}
\newcommand{\f}{\frac}

\title {On the
particle paths and the stagnation points in small-amplitude
deep-water waves}

\author{\normalsize Delia IONESCU-KRUSE\\
\normalsize Institute of Mathematics of
the Romanian Academy,\\
\normalsize P.O. Box 1-764, RO-014700, Bucharest,
 Romania\\
\normalsize E-mail: Delia.Ionescu@imar.ro\\[10pt]}

 \date{}

\begin{document}
\maketitle

\begin{abstract}
In order to obtain quite precise information about the shape of
the particle paths  below small-amplitude gravity waves travelling
on irrotational deep water,   analytic solutions of the nonlinear
differential equation system describing the particle motion are
provided. All these solutions are not closed curves. Some particle
trajectories are peakon-like, others can be expressed with the aid
of the Jacobi elliptic functions or with the aid of the
hyperelliptic functions.  Remarks on the stagnation points of the
small-amplitude irrotational deep-water waves are also made.
\end{abstract}

\section{Introduction}

The purpose of the present work is to investigate the fluid
dynamics at the propagation of small-amplitude gravity waves over
irrotational deep water. It was widely believed that as
small-amplitude gravity waves propagate on the surface, the
particles of the fluid move on circular orbits, the diameter of
which decreases with depth (see, for example, \cite{debnath},
\cite{johnson-carte},  \cite{lamb}, \cite{lighthill}). Indeed, in
the first approximation,  the fluid particles follow closed paths,
but,  by a phase-plane analysis of the nonlinear system describing
the particle motion, in  \cite{cv} and \cite{cev} it is shown that
no particle trajectory is actually closed, unless the free surface
is flat, there exists a forward drift over a period which
decreases with greater depth. Similar conclusions hold also within
the full nonlinear framework of the water-wave problem for
symmetric periodic steady gravity waves (Stokes waves) travelling
over a flat bed or over water of infinite depth.   For a
qualitative description of the particle trajectories in this
framework, by methods from the theory of harmonic functions, see
\cite{c2007} and \cite{henry} (see also the very recent book
\cite{const-carte}). The results in \cite{c2007} are recovered by
a simpler approach in \cite{cs2010}, where there are also
described all possible particle trajectories beneath a Stokes
wave. The particle trajectories change considerably according to
whether the Stokes waves enter a still region of water or whether
they interact with a favorable or adverse uniform current. Some
particle trajectories are closed orbits, some are undulating paths
and most are looping orbits that drift either to the right or to
the left, depending on the underlying current.\\
Notice that there are only a few explicit solutions to the full
nonlinear water-wave problems. For periodic gravity water waves in
water of infinite depth, Gerstner  constructed an explicit
solution\footnote{This solution was independently re-discovered
later by Rankine \cite{rankina}. Modern detailed descriptions of
this wave are given in the recent papers \cite{c2001a} and
\cite{henry4}.} in 1802 \cite{gerstner}. Gerstner's wave is a
two-dimensional wave given in the Lagrangian description, by
following  the evolution of individual water particles. The motion
of the water body induced by the passage of Gerstner's wave is
rotational, it occurs in a flow with a specific non-constant
vorticity. The fact that this flow is very special is confirmed
also by the fact that this is the only steady flow satisfying the
constraint of constant pressure along the streamlines cf.
\cite{kalisch}.  Beneath Gerstner's wave it is possible to have a
motion of the fluid where all particles describe circles with a
depth-dependent radius \cite{c2001a}, \cite{henry4}.

 In order to obtain quite precise information
about the shape of the particle paths  below small-amplitude
gravity waves travelling on irrotational deep water, in this paper
we provide analytic solutions of the nonlinear differential
equation system describing the particle motion. We show  that all
these solutions are not closed curves.
 In the study
 of the nonlinear  system (\ref{diff2}),
 a peakon-like trajectory  (\ref{sol0}) comes up. This solution has a vertical asymptote in the
positive direction (see Figure 2). A peakon-like solution appeared
also as particle path below small-amplitude periodic gravity waves
travelling  on a constant vorticity current (see \cite{io5} and
\cite{io6}).
 The other solutions of the nonlinear system (\ref{diff2}) are given
by (\ref{sol2}). They cannot be expressed in terms of elementary
functions. In some cases these solutions can be expressed with the
aid of the Jacobi elliptic functions, in other cases they can be
expressed with the aid of the hyperelliptic functions. We draw
some of the curves obtained for different values of the parameters
(see Figure 4, Figure 5, Figure 6). We observe that some solutions
(see (\ref{Z2})) have vertical asymptotes in the positive
direction (see Figure 6 too). This surprising feature is also
found at the solutions describing the motion of the particles
beneath small-amplitude capillary-gravity waves which propagate on
the surface of an irrotational water flow with a flat bottom (see
\cite{io4}). The particle seems to be shot out from the flow, this
feature  could reflect the wave-breaking phenomenon.\\
For additional information  on the particle paths within different
types of progressive water waves, in the framework of linear
theory or in the framework of full nonlinear theory of periodic
symmetric waves, and in the presence or not of the background
currents and vorticity, see the following references:
\cite{c2007}-\cite{cs2010}, \cite{henry}-\cite{henry3},
\cite{io}-\cite{io6}, \cite{matioc}, \cite{w2}.

Another remarkable feature that we investigate in this paper are
the stagnation points for our problem, that is,  points where the
vertical component of the fluid velocity field is zero while the
horizontal component equals the speed of the wave profile. The
stagnation points are of special interest because they are points
where the flow characteristic often change.  They could be located
on the free surface, in this case  the wave is called  extreme
wave, on the bottom or inside the fluid domain. For irrotational
flows, the existence of  extreme waves was predicted by  Stokes
\cite{stokes}, who also conjectured in 1880 that their profiles
necessarily have corners with an angle of 120$^o$ at the crest. A
rigorous proof of this conjecture has not been given until 1982,
when it was established
 independently in \cite{amick} and \cite{plotnikov}. For rotational flows,
the existence  theory of small-amplitude waves is due to
\cite{dubreil}, \cite{goyon}, \cite{zeidler}, while the existence
of rotational waves of large amplitude (approaching flows with
stagnation points)  in flows without stagnation points was
recently established in \cite{cs2004}, \cite{cs2011}. Some
considerations about the possible shape of the limiting flow of
the waves that almost admit a stagnation point are made in the
paper \cite{varvaruca1}. But in the rotational case, for
small-amplitude water waves with constant vorticity  the
stagnation points can also occur inside the fluid domain, as shown
in \cite{ev} and \cite{io6}. The papers \cite{w2} and \cite{cv2}
contain existence results for small-amplitude steady waves with
constant vorticity in the presence of stagnation points in the
flow. The stagnation points around which the fluid rotates are
centers. In irrotational deep water, there  were revealed in
\cite{luko},
 although numerical and
not completely rigorous (see \cite{const}), stagnation points
inside the fluid domain
 with discontinuous streamlines near the wave crests. These stagnation points are
  of saddle-point type.  The formation of the stagnation points inside the
flow may also be connected with a wave-breaking phenomenon in deep
water; the traditional criterion for wave breaking is that
horizontal  water velocities in the crest must exceed the speed of
the crest, see \cite{banner} where there are also reviewed
different  criterions, methods, measurements and
laboratory studies to detect breaking events in deep water.\\
Our solutions (\ref{sol0}) and (\ref{sol2}) are also used to
identify the stagnation points in small-amplitude irrotational
deep-water waves. For the solution (\ref{sol0}),  a stagnation
point in the fluid appears only for $t\rightarrow \pm\infty$. At
this point the path of the particle has a horizontal tangent and
the location of this stagnation point is on the bottom
$z=-\infty$. For the solution (\ref{sol2}), the stagnation points
are obtained by solving the equation (\ref{53}). This equation can
be solved graphically and depending on the signs and on the values
of the involved parameters it can have one,  two or three
solutions.
 Which of these
solutions are inside the fluid and their nature can be determined
by a further study.

\section{The general problem for propagation of gravity waves in
deep water} Deep-water waves are modelled mathematically as
periodic two-dimensional waves in water of infinite depth.  To
describe these waves we consider a cross section of the flow that
is perpendicular to the crest line with Cartesian coordinates
$(x,z)$, the $x$-axis being in the direction of wave propagation
and the $z$-axis pointing vertically upwards. The water flow under
consideration is bounded above by the free surface
 $z=\eta(x,t)$ and has the bottom at $z=-\infty$. The fluid is acted on only by the constant gravitational acceleration
$g$,  the effects of surface tension being ignored. For gravity
water waves, the appropriate equations of motion are Euler's
equations (see \cite{johnson-carte}):
\begin{equation}
\begin{array}{c}
u_t+uu_x+vu_z=-\f1{\rho} p_x\\
  v_t+uv_x+vv_z=-\f1{\rho} p_z-g,\\
\end{array}
  \label{e}
\ee where $(u(x,z,t), v(x,z,t))$ is the velocity field of the
water, $p(x,z,t)$ denotes the pressure,  and $\rho(x,z,t)$ is the
density. Another realistic assumption for gravity water wave
problem is the incompressibility, that is, the density $\rho$ is
constant (see \cite{lighthill}), which implies the equation of
mass conservation:
\begin{equation}
\begin{array}{c}
u_x+v_z=0.
  \end{array} \label{mc}
\ee
 The boundary conditions that define water-wave problems come in various forms.
 For our problem,
the dynamic boundary condition  expresses the fact that on the
free surface the pressure is equal to the constant atmospheric
pressure denoted $p_0$:
\begin{equation}
\begin{array}{c}
 p=p_0 \,  \textrm{ on } z=\eta(x,t). \end{array} \label{dbc}
 \end{equation}
The kinematic boundary condition  expresses the fact that the same
particles always form the free-water surface:
\begin{equation}
\begin{array}{c}
  v=\eta_t+u\eta_x \, \, \textrm{ on }\,
z=\eta(x,t),\end{array} \label{kbc}
 \end{equation}
 and the fact that at
 great depth there is practically no motion is expressed by the following boundary condition:
\begin{equation}
\begin{array}{c}
 (u, v)\rightarrow (0,0)  \, \, \textrm{ as }\, z\rightarrow
 -\infty.\end{array} \label{bc}
 \end{equation}
The governing equations and the boundary conditions
(\ref{e})-(\ref{bc}) define the boundary-value problem for
deep-water waves.

 An important category
of flows is those of zero vorticity (irrotational flows),
characterized by the additional equation: \be u_z-v_x=0.
\label{vorticity}\ee In what follows we will consider this type of
flow. The idealization of irrotational flow is physically relevant
in the absence of non-uniform currents in the water.

For the water wave problem, there was a great interest in the
regularity, or  even stronger, the real-analyticity,  of the
solutions and particularly of the free surface. If these functions
are real-analytic in some region, we can approximate them by using
power series, approach used in small-amplitude wave theory. In the
irrotational case, Lewy's theorem \cite{lewy} shows that the
free-surface must be a real-analytic curve and that the complex
potential must have an analytic extension across the boundary (see
\cite{toland}). Recent developments providing the regularity/real
analyticity results for rotational flows were initiated in
\cite{CE3} for gravity water waves over a flat bed; for deep-water
waves see \cite{bmatioc}. In the case of zero vorticity, the
approach in  \cite{CE3} yields an alternative short proof of the
 the famous results obtained previously by Lewy.
\\

 \hspace{2cm}\scalebox{0.65}{\includegraphics{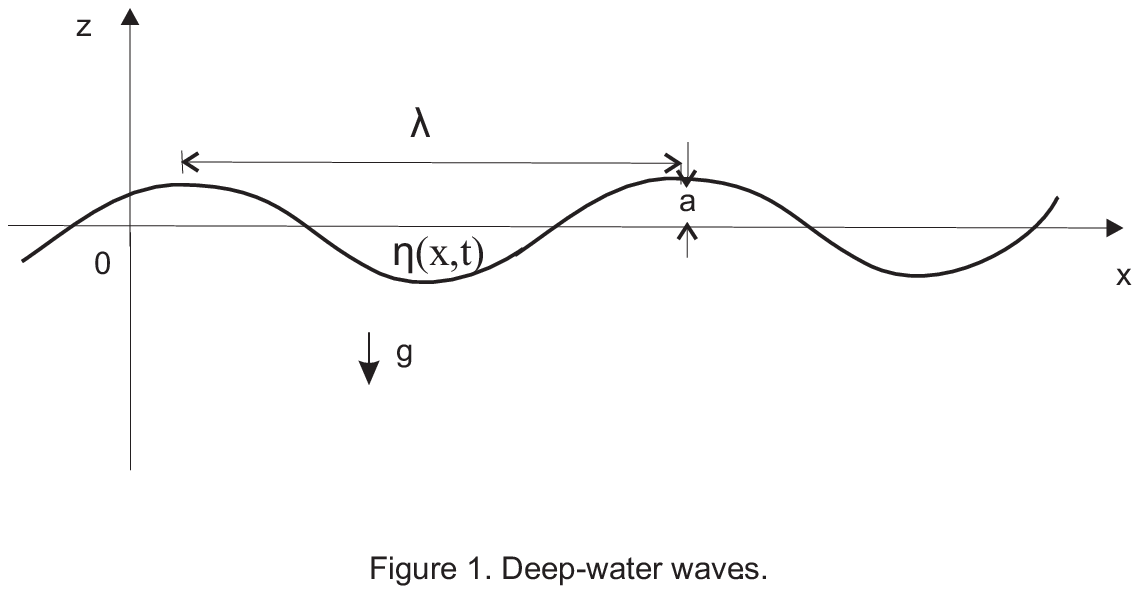}}
\\

\section{Small-amplitude approximation of the
  water-wave problem}

Let us non-dimensionalise now the set of equations
(\ref{e})-(\ref{vorticity}). If $\lambda>0$ is the wavelength and
$a>0$ is the wave amplitude (see Figure 1.), we make the following
change of variables (for more details see \cite{johnson-carte}),
\begin{equation}
\begin{array}{c}
x\mapsto\lambda x,   \quad z\mapsto z, \quad  \eta\mapsto a\eta,
\quad t\mapsto\f\lambda{\sqrt{g}}t,\\
  u\mapsto  \sqrt{g}u,
\quad v\mapsto \f{\sqrt{g}}{\lambda}v,
\end{array} \label{nondim}\end{equation}
\begin{equation} p\mapsto p_0-gz+ g  p,
\label{p}\end{equation} where, to avoid new notation, we have used
the same symbols for the non-dimensional variables  $x$, $z$,
$\eta$, $t$, $u$, $v$, $p$ on the right-hand side. We set the
constant water density $\rho=1$.
 Therefore, in
non-dimensional variables (\ref{nondim}), (\ref{p}), the
water-wave problem (\ref{e})-(\ref{vorticity}) becomes:
\begin{equation}
\begin{array}{cc}
u_t+uu_x+vu_z=- p_x&\\  \f 1{\lambda^2}(v_t+uv_x+vv_z)=- p_z&\\
 u_x+v_z=0&\\
  u_z-\f 1{\lambda^2}v_x=0&\\
  v=a(\eta_t+ u\eta_x)  \, & \textrm{ on }\,
z=a\eta(x,t)\\
p=a\eta \, & \textrm{ on }\,
z=a\eta(x,t)\\
(u, v)\rightarrow (0,0)  \, &  \textrm{ as }\, z\rightarrow
 -\infty.
 \end{array}
\label{e+bc'} \end{equation} We observe  in (\ref{e+bc'}) that on
$z=a\eta$ both $v$ and $p$ are proportional to $a$, this being
consistent with the fact that as $a\rightarrow 0$ we must have
$v\rightarrow 0$ and $p\rightarrow 0$ (with no disturbance the
free surface becomes a horizontal surface on which $v=p=0$). For
consistence, one requires that $u$ is proportional to $a$ too.
Thus, by the following scaling of the non-dimensional variables:
\begin{equation} p\mapsto a p,\quad
u\mapsto a u,\quad v\mapsto a v, \label{scaling}\end{equation}
where we avoided again the introduction of a new notation, the
problem (\ref{e+bc'}) becomes:
\begin{equation}
\begin{array}{cc}
u_t+a(uu_x+vu_z)=- p_x&\\  \f 1 {\lambda^2}[v_t+a(uv_x+vv_z)]=- p_z&\\
 u_x+v_z=0&\\
  u_z-\f 1{\lambda^2}v_x=0&\\
   v=\eta_t+ a u\eta_x  \, & \textrm{ on }\,
z=a\eta(x,t)\\
p=\eta \, & \textrm{ on }\,
z=a\eta(x,t)\\
(u, v)\rightarrow (0,0)  \, &  \textrm{ as }\, z\rightarrow
 -\infty.
 \end{array}
\label{e+bc1''} \end{equation} The classical approximation is the
linearized problem obtained by requiring the amplitude of the free
surface to be small, that is, by letting $a\rightarrow 0$,
$\lambda$ being fixed. Thus, we obtain the linear system:
% we
%get the following linear system
\begin{equation}
\begin{array}{cc}
u_t+p_x=0&\\ \f 1{\lambda^2}v_t+ p_z=0&\\
 u_x+v_z=0&\\
 u_z-\f 1{\lambda^2}v_x=0&\\
v=\eta_t  \, & \textrm{ on }\,
z=0\\
  p=\eta \, & \textrm{ on }\,
z=0\\
(u, v)\rightarrow (0,0)  \, &  \textrm{ as }\, z\rightarrow
 -\infty.
\end{array}
\label{small} \end{equation}
 From the third equation and the forth  equation in
(\ref{small}), we get that \be v_{zz}+\f
1{\lambda^2}v_{xx}=0.\label{17'}\ee Applying the method of
separation of variables, we seek the solution of the  equation
(\ref{17'}) in the form \be v(x,z,t)=F(x,t)G(z,t). \label{19}\ee
Substituting (\ref{19}) into the equation (\ref{17'}), we find \be
F\f{\pa^2 G}{\pa z^2}+\f 1{\lambda^2} G\f{\pa^2F}{\pa x^2}=0, \ee
thus, \be\f{1}{G}\f{\pa^2 G}{\pa z^2}=-\f
1{\lambda^2}\f{1}{F}\f{\pa^2 F}{\pa x^2}.\ee We observe in the
above equation that the left hand side does not depend on $z$ and
the right hand side does not depend on $x$. Therefore, each side
must be a constant, say \be \f{1}{F}\f{\pa^2 F}{\pa
x^2}=-K^2,\quad \f{1}{G}\f{\pa^2 G}{\pa z^2}=\f{K^2}{\lambda^2}
\label{20} \ee where $K\geq 0 $ is a constant that might depend on
time. With the above choice, the solutions of the equations in
(\ref{20}) are \ba
F(x,t)&=&\mathcal{A}\sin(Kx)+\mathcal{B}\cos(Kx)\nonumber\\
G(z,t)&=&\mathcal{C}e^{\f K{\lambda} z}+\mathcal{D}e^{-\f
K{\lambda} z},\nonumber \ea where $\mathcal{A}$, $\mathcal{B}$,
$\mathcal{C}$, $\mathcal{D}$ are constants depending on time. From
the last condition in (\ref{small}), that is, $v\rightarrow 0$ as
$z\rightarrow -\infty$,  we get that $\mathcal{D}=0$, and thus,
\be v(x,z,t)=\mathcal{C}e^{\f K{\lambda}
z}[\mathcal{A}\sin(Kx)+\mathcal{B}\cos(Kx)].\ee On $z=0$, by the
fifth equation in (\ref{small}), we have $v=\eta_t$, which yields
\be
\mathcal{C}[\mathcal{A}\sin(Kx)+\mathcal{B}\cos(Kx)]=\eta_t.\ee
 Therefore,  \be v(x,z,t)=e^{\f K{\lambda} z} \eta_t.
\label{43}\ee Taking into account (\ref{43}) and the forth
equation in (\ref{small}), we get \be u(x,z,t)=\f 1{K\lambda}
e^{\f K{\lambda} z}\eta_{tx}+\mathcal{F}(x,t),\label{44}\ee
$\mathcal{F}(x,t)$ being an arbitrary function. The last condition
in (\ref{small}), that is, $u\rightarrow 0$ as $z\rightarrow
-\infty$, yields \be \mathcal{F}(x,t)=0.\label{45}\ee
 The components $u$
and $v$ of the velocity have to fulfill also the third equation in
(\ref{small}). Replacing (\ref{43}), (\ref{44}) with (\ref{45}),
in this equation, we obtain the equation that $\eta$ has to
fulfill: \be \eta_{txx}+K^2\eta_t=0. \label{25}\ee
 We
seek periodic travelling wave solutions; thus, for the equation
(\ref{25}) with \be K:=2\pi,\label{46}\ee
  we choose the following solution \be
\eta(x,t):=\cos(2\pi(x-ct)), \label{26}\ee where $c$ represents
the non-dimensional speed of propagation of the linear wave and is
to be determined.  With (\ref{46}) and (\ref{26}) in view, the
components of the velocity field become \ba && u(x,z,t)=\f
{2\pi c}{\lambda} e^{\f {2\pi}{\lambda} z}\cos(2\pi(x-ct))  \nonumber\\
&&v(x,z,t)=2\pi c e^{\f {2\pi}{\lambda} z}\sin(2\pi(x-ct)).    \ea
In order to find the expressions of the pressure we take into
account the first two equations in (\ref{small}) and the
expressions of the velocity field from above. Thus, we obtain  \be
p(x,z,t)=\f{2\pi c^2}{\lambda}e^{\f {2\pi}{\lambda}
z}\cos(2\pi(x-ct))+c_0 \label{29}\ee where $c_0$ is a constant. On
$z=0$ the pressure (\ref{29}) has to fulfill the sixth equation of
system (\ref{small}). Hence, in view of (\ref{26}), we get \be
\f{2\pi c^2}{\lambda}e^{\f {2\pi}{\lambda}
z}\cos(2\pi(x-ct))+c_0=\cos(2\pi(x-ct)). \ee The above relation
must hold for all values $x\in \mathbf{R}$; therefore, we obtain
\be c_0=0,\ee and we provide the non-dimensional speed of the
linear wave \be c^2=\f{\lambda}{2\pi}.\label{c}\ee We observe that
the speed of the linear wave is proportional to the square root of
the wavelength. In other words, long water waves travel faster
than shorter waves. \\
Summing up, the system (\ref{small}) has the
solution \be
 \begin{array}{llll}
 \eta(x,t)=\cos(2\pi(x-ct))\\
u(x,z,t)=\f
{2\pi c}{\lambda} e^{\f {2\pi}{\lambda} z}\cos(2\pi(x-ct))\\
v(x,z,t)=2\pi c e^{\f {2\pi}{\lambda} z}\sin(2\pi(x-ct))\\
 p(x,z,t)=e^{\f {2\pi}{\lambda}
z}\cos(2\pi(x-ct)),
\end{array}\label{solrot0}\ee with $c$ given by
(\ref{c}).
% This solution is also obtained in \cite{cev} (see 2.9).

Taking into account (\ref{nondim}), (\ref{p}), (\ref{scaling}), we
return to the original physical variables. The speed of the wave
(\ref{c}) and the solution (\ref{solrot0}) become: \be
c=\pm\sqrt{g}\sqrt{\f{\lambda}{2\pi}}=\pm \sqrt{\f
{g}{k}}\label{c1} \ee \be
 \hspace{0cm}\begin{array}{llll}
  \eta(x,t)=
  a\cos(k(x-ct))\\\cr
 p(x,z,t)=p_0-gz+age^{kz} \cos(k(x-ct))\\\cr u(x,z,t)=acke^{kz}\cos(k(x-ct))\\
\cr v(x,z,t)=acke^{k z}\sin(k(x-ct)),
  \end{array}\label{solrotconst'}\ee
where \be k:=\f{2\pi}{\lambda}\ee is the wave number. The sign
minus in (\ref{c1}) indicates a left-going wave. We observe that
the velocity components have the same amplitude $acke^{kz}$ which
depends on position and decreases exponentially with the distance
 below the surface.

\section{Particle
trajectories}

Let $\left(x(t), z(t)\right)$ be the path of a particle in the
fluid domain, with location $\left(x(0), z(0)\right):=(x_0,z_0)$
at time $t=0$. The motion of the particles below the
 small-amplitude  deep-water waves  with the velocity field  (\ref{solrotconst'}), is described
by the following differential system:
  \be\left\{\begin{array}{ll}
 \f{dx}{dt}=u(x,z,t)=Ae^{kz}\cos(k(x-ct))\\
\\
 \f{dz}{dt}=v(x,z,t)=Ae^{kz}\sin(k(x-ct)),
 \end{array}\right.\label{diff2}\ee
where the constant $A$ is \be A:=ack\neq 0.\label{a} \ee The sign
of $A$ depends  on the sign of the wave speed $c$. Thus, if we
choose in (\ref{c1}) the square root with minus, that is, we
consider left-going waves, we have $A<0$ and if we choose in
(\ref{c1}) the square root with plus, that is, we consider
right-going waves, we get $A>0$.

 To study
the exact solution of the system (\ref{diff2}) it is
 more convenient to re-write it in the following moving frame
 \be
 X=k(x-ct),\quad  Z=k z. \label{frame}
 \ee
This transformation yields \be\left\{\begin{array}{ll}
 \f{dX}{dt}=kAe^Z\cos(X)-kc\\
 \\
 \f{dZ}{dt}=kA e^Z\sin(X).
 \end{array}\right.\label{diff3}\ee
We write the second equation of this system in the form \be
e^{-Z}dZ=kA\sin X(t)\,dt. \label{47}\ee Integrating, we get \be
-e^{-Z}=\int kA \sin X(t)\,dt.\label{X} \ee If \be \int kA \sin
X(t)\,dt<0, \label{int}\ee then, \be Z(t)=-\log\left[-\int kA \sin
X(t)\,dt\right]. \label{42}\ee We  denote by \be w=w(t):=\int k A
\sin X(t)\,dt. \label{31}\ee With (\ref{int}) in view, we have \be
w<0.\label{w} \ee
 From (\ref{31}) we get \be
kA\sin X(t)=\f{d w}{dt}. \label{32}\ee Differentiating with
respect to $t$ this relation, we obtain
 \be k A\cos (X)\f{dX}{dt}=\f{d^2w}{dt^2}.\label{33}
\ee From (\ref{32}) we have furthermore: \be k^2A^2\cos^2
(X)=k^2A^2-\left(\f{d w}{dt}\right)^2.\label{34} \ee Thus, taking
into account (\ref{42}), (\ref{31}), (\ref{33}) and (\ref{34}),
the first equation of the system  (\ref{diff3}) becomes
\begin{eqnarray} &&\f{d^2w}{dt^2}=\left(-\f{1}{w}\right)\left[k^2A^2-\left(\f{dw}{dt}
 \right)^2\right]-kc\sqrt{k^2A^2-\left(\f{dw}{dt}
 \right)^2}.\label{36'}
\end{eqnarray}
We make the following substitution
 \be \xi^2(w):=k^2A^2-\left(\f{dw}{dt}\right)^2, \label{37}\ee
$A$ being different from zero by (\ref{a}).
 Differentiating with respect to $t$ this relation, we get
 \be
\xi\f{d\xi}{dw}=-\f{d^2w}{dt^2}.
 \label{38}\ee
We replace (\ref{37}), (\ref{38}) into the equation (\ref{36'})
and we obtain  the equation \be
\xi\f{d\xi}{dw}=\f{1}{w}\xi^2+kc\xi. \label{39c}\ee A solution of
the equation (\ref{39c}) is \be \xi=0,\ee which, in view of
(\ref{37}) and (\ref{32}) implies \be \sin X(t)=\pm 1. \ee
Therefore, from (\ref{42}) with the condition (\ref{int}),  and
further from (\ref{frame}), \textit{a solution of the system
(\ref{diff2})  is \be
\begin{array}{ll}
 x(t)=ct+\textrm{const}_1\\
 \\
 z(t)=-\f 1{k}\log\left(|kA\,t+\textrm{const}_2|\right),
  \end{array} \label{sol0}\ee}
$const_1$ and $const_2$ are constants determined by the initial
conditions $\left(x(0), z(0)\right):=(x_0,z_0)$.
 We observe that
\ba &&\lim_{t\rightarrow
-\f{\textrm{const}_2}{kA}}x(t)=\textrm{const},\nonumber\\
&&\\ &&\lim_{\substack{t\rightarrow
-\f{\textrm{const}_2}{kA}\\t>-\f{\textrm{const}_2}{kA}
}}z(t)=\lim_{\substack{t\rightarrow
-\f{\textrm{const}_2}{kA}\\t<-\f{\textrm{const}_2}{kA}
}}z(t)=+\infty, \nonumber\ea and \be \lim_{t\rightarrow
\pm\infty}x(t)=\pm\infty,\quad \lim_{t\rightarrow
\pm\infty}z(t)=-\infty. \ee Therefore,  $x=\textrm{const}$ will be
a vertical asymptote and $z=-\infty$ will be a horizontal
asymptote for  the curve (\ref{sol0}). The particle seems to be
shot out from the flow, this feature  could reflect the
wave-breaking phenomenon. For $c>0$,  the graph of the parametric
curve (\ref{sol0}) is drawn  in the
  Figure 2. For $c<0$, the arrows will change the direction.
 This peakon-like solution appeared also as particle path below
small-amplitude periodic gravity waves travelling  on  a constant
vorticity current (see \cite{io5} and \cite{io6}).

\vspace{0.6cm}

\hspace{0.5cm} \scalebox{0.40}{\includegraphics{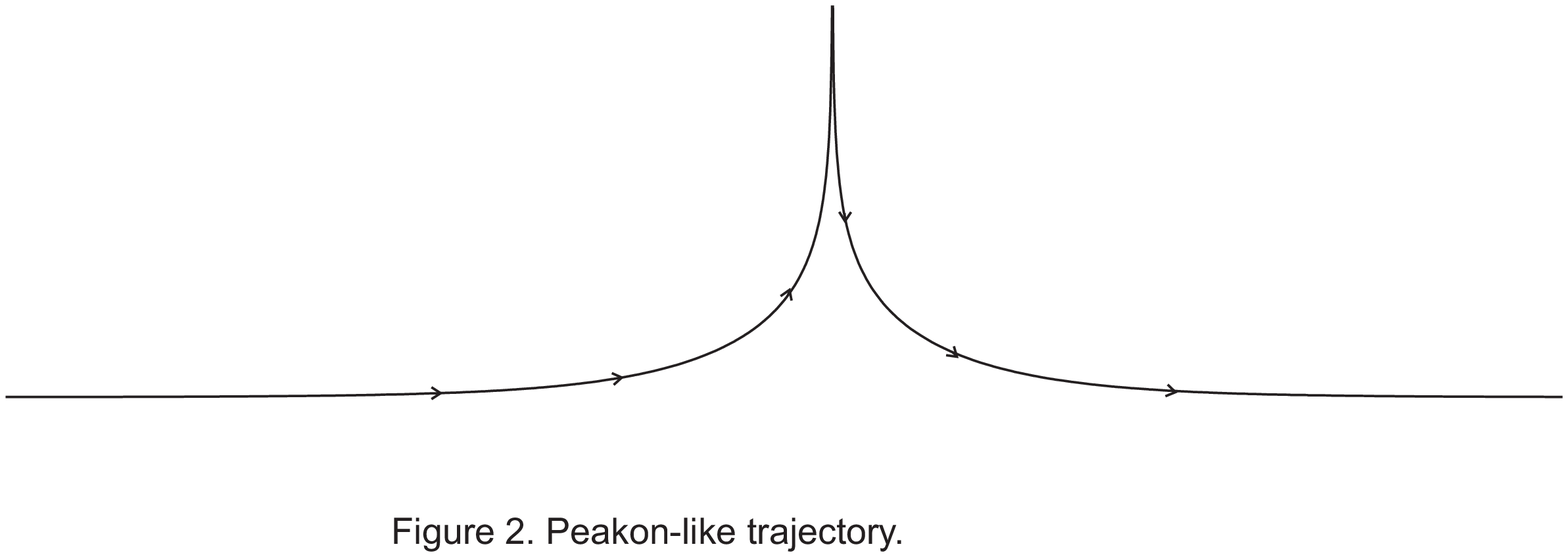}}
\\

\noindent Calculating the derivatives of $x(t)$ and $z(t)$ with
respect to $t$, we get
 \be
\begin{array}{ll}
x'(t)=c\\
\cr z'(t)=-\f{A}{kA\,t+const_2}.
\end{array}\ee
Hence, \textit{for the solution (\ref{sol0}) a stagnation point in
the fluid,  where  $x'(t)=c$, $z'(t)=0$, appears only for
$t\rightarrow \pm\infty$. We observe that at this point the path
of the particle has a horizontal tangent and  is located on the
bottom $z=-\infty$}.

The other solutions of the equation (\ref{39c}) satisfy
 \be
\f{d\xi}{dw}=\f{\xi}{w}+ kc. \label{39c'} \ee The homogeneous
equation \be \f{d\xi}{\xi}=\f{dw}{w}\ee has the solution \be
\xi(w)= \mathcal{\theta}w,\ee where $\theta$ is an integration
constant. By the method of variation of constants, the general
solution of the non-homogeneous equation (\ref{39c'}) is given by
\be \xi(w)=\theta(w)w, \ee where $\theta(w)$ is a continuous
function which satisfies the equation \be
\f{d\theta}{dw}=\f{kc}{w}. \label{40c}\ee With (\ref{w}) in view,
the solution of the equation (\ref{40c}) is \be \theta(w)=
kc\log(-w)+\beta,\ee $\beta$ being a constant. Therefore, the
solution of the non-homogeneous equation (\ref{39c'}) has the
expression \be \xi(w)=w\left[kc\log(-w)+\beta\right]. \ee Further,
taking into account (\ref{37}), we get the equation that $w(t)$
has to fulfill: \be
\left(\f{dw}{dt}\right)^2=k^2A^2-w^2\left[kc\log(-w)+\beta\right]^2.\label{eq}
\ee
% Solving  this equation we would get $w=w(t)$ and then, by
%(\ref{31}), (\ref{42}), we would obtain $X(t)$ and $Z(t)$.
From (\ref{X}) and  (\ref{31}),  between $w(t)$ and $Z(t)$ there
exists the following relation:
 \be w(t)=-e^{-Z(t)}.\ee Therefore, we
obtain from (\ref{eq}) the differential equation that $Z(t)$ has
to satisfy: \be \left(\f{dZ}{dt}\right)^2=
k^2A^2e^{2Z}-\left[kcZ-\beta\right]^2.\label{48}\ee By separating
the variables, we get \be
\pm\f{dZ}{\sqrt{k^2A^2e^{2Z}-\left[kcZ-\beta\right]^2}}=dt.\label{49}
\ee Thus, taking into account  (\ref{47}) and (\ref{frame}),
\textit{the other solution of the system (\ref{diff2}) has the
following expression: \be
\begin{array}{llll}
 x(t)= ct+\f{1}{k}\arcsin \left[\f 1{kA}e^{-Z(t)}\f{dZ(t)}{dt}\right]\\
 \hspace{0.6cm}\stackrel{(\ref{49})}{=}
 ct\pm\f{1}{k}\arcsin \left[\sqrt{1-\left[\f{kcZ(t)-\beta}{kAe^{Z(t)}}\right]^2}\right]\\
%\hspace{0.7cm} =ct\pm\f{1}{k}\arcsin \left[
%\sqrt{1-\left[\f{2\beta-k(C-c)Z-\f{B}{2}
%Z^2}{kA\sinh(Z(t))}\right]^2}\right]\\
\\
 z(t)=\f 1{k} Z(t)
  \end{array} \label{sol2}\ee
 $Z(t)$ being the solution of the equation (\ref{49})}. The
 constant $\beta$ is determined by the initial conditions $\left(x(0), z(0)\right):=(x_0,z_0)$.

We observe that the solutions (\ref{sol2}) \textit{are not closed
curves}.

\noindent Indeed, if there exists $t_2>t_1$ such that
$Z(t_2)=Z(t_1)$, then,  we get
 $z(t_2)=z(t_1)$ and
$x(t_2)-x(t_1)
%=c(t_2-t_1)+\text{\tiny{$\f{1}{k}\arcsin \left[
%\sqrt{1-\left[\f{2\beta-k(C-c)Z(t_2)-\f{B}{2}
%Z^2(t_2)}{kA\sinh(Z(t_2))}\right]^2}\right]-\f 1{k}\arcsin \left[
%\sqrt{1-\left[\f{2\beta-k(C-c)Z(t_1)-\f{B}{2}
%Z^2(t_1)}{kA\sinh(Z(t_1))}\right]^2}\right]$}}
=c(t_2-t_1)\neq 0$. If $c>0$, the particles which follow these
curves will have a \textit{forward drift}, if $c<0$, they will
have a \textit{backward drift}.$\square$
\\

Let us make some remarks on the stagnation points inside the
fluid. Calculating the derivatives  with respect to $t$ of $x(t)$
and $z(t)$ from (\ref{sol2}), we get
 \be
\begin{array}{ll}
x'(t)=c+\f1{k}\f{\f{d^2Z}{dt^2}-\left(\f{dZ}{dt}\right)^2}{\sqrt{k^2A^2e^{2Z}
-\left(\f{dZ}{dt}\right)^2}}\\
\cr z'(t)=\f 1{k}\f{dZ}{dt}.\end{array}\label{52}\ee
 With
(\ref{49}) in view, for those $Z$ satisfying the following
equation \be \Big| kAe^Z\Big|=\Big|kcZ-\beta\Big| ,\label{53}\ee
%and its derivative, that is, the
%equation \be kA\cosh(Z)=-k(C-c)-BZ,\ee
we have \be \f{dZ}{dt}=0,\quad \f{d^2Z}{dt^2}=0,\ee
%\ba&&\f{dZ}{dt}=0,\nonumber\\
%&&\f{d^2Z}{dt^2}=kA\sinh(Z)\left[kA\cosh(Z)+k(C-c)+B
%Z\right]=0,\ea
 and thus, $x'(t)$, $z'(t)$  from
 (\ref{52}) becomes
 \be
 x'(t)=c,\quad z'(t)=0.
 \ee
Hence, \textit{for the solution (\ref{sol2}), the stagnation
points in the fluid are obtained by solving the equation
(\ref{53})}.

The equation (\ref{53}) can be solved graphically. Depending on
the signs and on the values of the parameters k, c, A and $\beta$,
the equation (\ref{53})  can have one,  two or three solutions.
 See, for
example, in Figure 3 some possibilities that can occur. With
continuous line we have drawn $\Big| kAe^Z\Big|$. Which of these
solutions are inside the fluid and their nature can be determined
by a further study.
\\

\hspace{-0.3cm} \scalebox{0.40}{\includegraphics{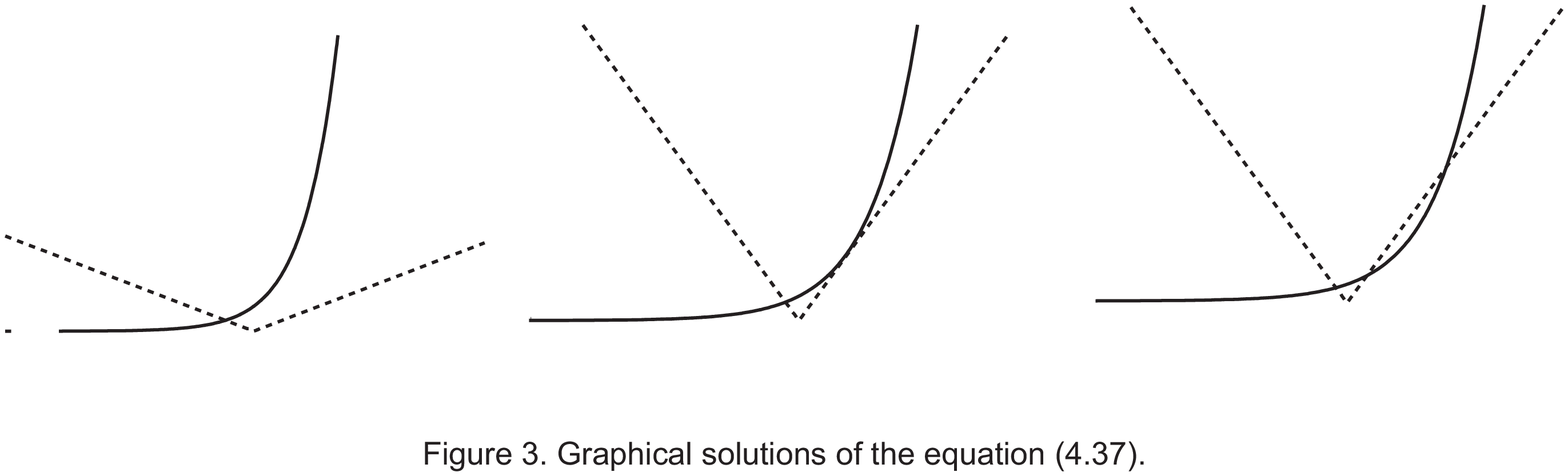}}
\\

Let us now investigate more the equation (\ref{49}). Taking into
account the expression of  $e^{2Z}$ as Taylor series: \be
e^{2Z}=\sum^\infty_{n=0}\f{(2Z)^{n}}{(n)!}=1+2Z+2Z^2+\f{4Z^3}{3}+\f{2Z^4}{3}+\f{4Z^5}{15}+\f{4Z^6}{45}+\cdots
\label{exp}\ee we get under the square root in (\ref{49}) the
following power series \ba &&(k^2A^2-\beta^2)+2k(kA^2+\beta c) Z
+k^2(2A^2-c^2)Z^2+\nonumber\\\cr
 &&\hspace{0.7cm}
+\f{4k^2A^2Z^3}{3}+\f{2k^2A^2Z^4}{3}+\f{4k^2A^2Z^5}{15}+\f{4k^2A^2Z^6}{45}+\cdots
\label{50}\ea The constant $A$ being different from zero
(\ref{a}),   the power series (\ref{50}) contains for sure powers
of $Z$ higher than two. A partial sum of the above series is a
polynomial of degree higher than two, therefore, in general,
$Z(t)$ cannot be expressed in terms of elementary functions.  If
in a partial sum of this series we consider powers of $Z$ smaller
or equal than four, then,
 the solution of the equation
(\ref{49}) involves an \textit{elliptic integral} (for elliptic
integrals see, for example, \cite{byrd}). Its inversion would lead
to an \textit{elliptic function}. If in a partial sum of the
series (\ref{50}) we consider powers of $Z$  higher than four,
then,
 the solution of the equation
(\ref{49}) involves a \textit{hyperelliptic integral} (for
hyperelliptic integrals see, for example, \cite{byrd}, page 252).
Its inversion would lead to a \textit{hyperelliptic function}.

If in (\ref{50})  we consider powers of $Z$ till three, then, the
solution of the equation (\ref{49}) involves the following
elliptic integral  of the first kind:
 \be
\pm\int\f{dZ}{\sqrt{\f{4k^2A^2}{3}Z^3+k^2(2A^2-c^2)Z^2+2k(kA^2+\beta
c) Z+(k^2A^2-\beta^2)}}=t.\label{50'} \ee This elliptic integral
of the first kind   may by reduced to the Legendre normal form.

\textbf{Case 1}: all the zeroes of the cubic polynomial under the
square root in (\ref{50'}) are real and distinct. We denote them
by $Z_1<Z_2<Z_3$. We introduce the variable $\varphi$ by (see
\cite{smirnov} Ch. VI, \S 4, page 602) \be
Z=Z_2\sin^2\varphi+Z_1\cos^2\varphi, \label{varphi}\ee and we get
\ba &&
\f{4k^2A^2}{3}(Z-Z_1)(Z-Z_2)(Z-Z_3)=\nonumber\\
&&\hspace{0.5cm}=\f{4k^2A^2}{3}
\sin^2\varphi\cos^2\varphi(Z_2-Z_1)^2(Z_3-Z_1)\left(
1-k_1^2\sin^2\varphi\right)>0\nonumber\\
&&dZ=2\sin\varphi\cos\varphi(Z_2-Z_1) d\varphi,\nonumber\ea where
the constant $0<k_1^2<1$ is given by \be
k_1^2:=\f{Z_2-Z_1}{Z_3-Z_1}.\label{k1}\ee Therefore we obtain the
Legendre normal form of the integral in (\ref{50'}):\be
\f{1}{\mathcal{C}_1}\int\f{d\varphi}{\sqrt{1-k_1^2\sin^2\varphi
}}=t, \label{51}\ee the constant factor in front of the integral
being equal to\be \mathcal{C}_1:=\pm\f
1{\sqrt{3}}k|A|\sqrt{Z_3-Z_1}.\label{C}\ee
 The inverse of the integral in (\ref{51}) is the Jacobi elliptic function sine amplitude
 sn
  (see, for example,
\cite{byrd})  \be \textrm{ sn }\left(\mathcal{C}_1\,t;k_1\right)
:=\sin\varphi.\label{56}\ee In view of the notation
(\ref{varphi}), we get \be Z(t)=Z_2\textrm{ sn
}^2\left(\mathcal{C}_1\,t;k_1\right)+Z_1\textrm{ cn
}^2\left(\mathcal{C}_1\,t;k_1\right), \label{Z1}\ee
 cn being the
Jacobi elliptic function cosine amplitude (see, for example,
\cite{byrd}). We introduce  (\ref{Z1})  in (\ref{sol2}) and we get
$x(t)$ and $z(t)$ explicitly.

\textbf{Case 2}: the cubic polynomial under the square root in
(\ref{50'}) has only one real solution denoted $Z_0$.
 We denote by $p$ and $q$ the real coefficients such that
\ba && \f{4k^2A^2}{3}Z^3+k^2(2A^2-c^2)Z^2+2k(kA^2+\beta c)
Z+(k^2A^2-\beta^2)=\nonumber\\
&&\hspace{1.5cm}=\f{4k^2A^2}{3}(Z-Z_0)(Z^2+pZ+q).\label{60}\ea We
introduce the variable $\psi$ by (see \cite{smirnov} Ch. VI, \S 4,
page 602) \be Z=Z_0+\sqrt{Z_0^2+pZ_0+q}\,\tan^2\f{\psi}{2},
\label{varphi2}\ee and we get \ba &&
\f{4k^2A^2}{3}(Z-Z_0)(Z^2+pZ+q)=\nonumber\\
&&\hspace{1.5cm}=\f{4k^2A^2}{3}
\left(\sqrt{Z_0^2+pZ_0+q}\right)^3\f{\tan^2\f{\psi}{2}}{\cos^4\f{\psi}{2}}\left(
1-k_2^2\sin^2\varphi\right)>0\nonumber\\
&&dZ=\sqrt{Z_0^2+pZ_0+q}\f{\tan\f{\psi}{2}}{\cos^2\f{\psi}{2}}
d\psi,\nonumber\ea where the constant $0<k_2^2<1$ is given by \be
k^2_2:=\f
1{2}\left(1-\f{Z_0+\f{p}{2}}{\sqrt{Z_0^2+pZ_0+q}}\right).\label{k2}
\ee Therefore we obtain the Legendre normal form of the integral
in (\ref{50'}): \be
\f{1}{\mathcal{C}_2}\int\f{d\psi}{\sqrt{1-k_2^2\sin^2\psi }}=t,
\label{51'}\ee the constant factor in front of the integral being
equal to \be
\mathcal{C}_2:=\pm\f{2}{\sqrt{3}}k|A|(Z_0^2+pZ_0+q)^\f1{4}.\label{C2}\ee
 The inverse of the integral in (\ref{51'}) is
   \be \textrm{ sn }\left(\mathcal{C}_2\,t;k_2\right)
=\sin\psi\label{56'}\ee Taking into account (\ref{varphi2}), we
get \be Z(t)=Z_0+\sqrt{Z_0^2+pZ_0+q}\,\f{1-\textrm{ cn
}\left(\mathcal{C}_2\,t;k_2\right)}{1+\textrm{ cn
}\left(\mathcal{C}_2\,t;k_2\right)} \label{Z2}\ee We introduce
(\ref{Z2})  in (\ref{sol2}) and we get $x(t)$ and $z(t)$
explicitly. \\
We observe that  for that $t$'s, denoted
$\tilde{t}+\mathcal{K}$, with $\mathcal{K}$ a period, for which
the periodic Jacobi elliptic function cn satisfies the equation
 \be
 1+\textrm{ cn
}\left(\mathcal{C}_2\,t;k_2\right)=0,\ee  we have,  in view of
(\ref{Z2})  and (\ref{sol2}), that \be \lim_{t\rightarrow
(\tilde{t}+K)}x(t)=\textrm{ const }:=\tilde{x}+K,\quad
\lim_{t\rightarrow (\tilde{t}+K)}z(t)=\infty \label{limit'}\ee
Therefore, $x=\tilde{x}+K$ will be vertical asymptotes in the
positive direction.
 This surprising feature is also found at the
solutions describing the motion of the particles beneath
small-amplitude capillary-gravity waves which propagate on the
surface of an irrotational water flow with a flat bottom (see
\cite{io4}). The particle seems to be shot out from the flow, this
feature  could reflect the wave-breaking phenomenon.\\

Let us draw below some of the curves  obtained for different
values of the parameters, using  Mathematica\footnote{In
Mathematica the Jacobi elliptic functions  are implemented  as
JacobiSN$[u,m:=k^2_1]$:=sn$(u;k_1)$,
JacobiCN$[u,m:=k^2_1]$:=cn$(u;k_1)$,
JacobiDN$[u,m:=k^2_1]$:=dn$(u;k_1)$}.

 We consider  $k=1$, $g=9.8$,
$a=0.1$ and  $\beta=1$. Then, by (\ref{c1}),  choosing the square
root with sign plus,  and by (\ref{a}), we get $c=3.1305$,
$A=0.31305$. In this case, all the roots of the cubic polynomial
under the square root in (\ref{50'}) are real and we get $Z(t)$ in
the form (\ref{Z1}). The graph of the curve obtained is drawn in
Figure 4.
\\

 \scalebox{0.40}{\includegraphics{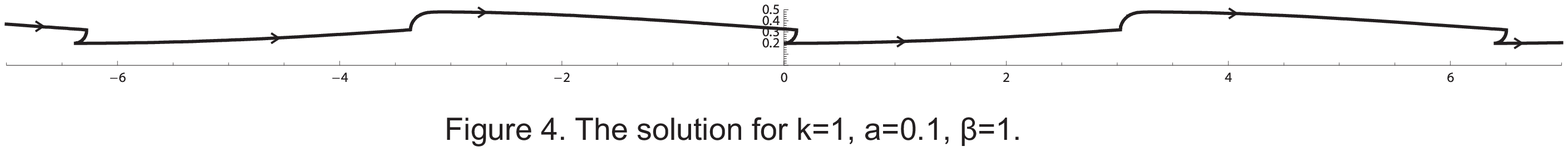}}
\\

For $k=2$,  $g=9.8$, $a=0.1$ and  $\beta=-1$, choosing the square
root with sign plus in (\ref{c1}), we get $c=2.21359$,
$A=0.442718$, all the roots of the cubic polynomial under the
square root in (\ref{50'}) are real, and thus,  $Z(t)$ has the
form (\ref{Z1}). The graph of the curve obtained is presented in
Figure 5.
\\

 \scalebox{0.40}{\includegraphics{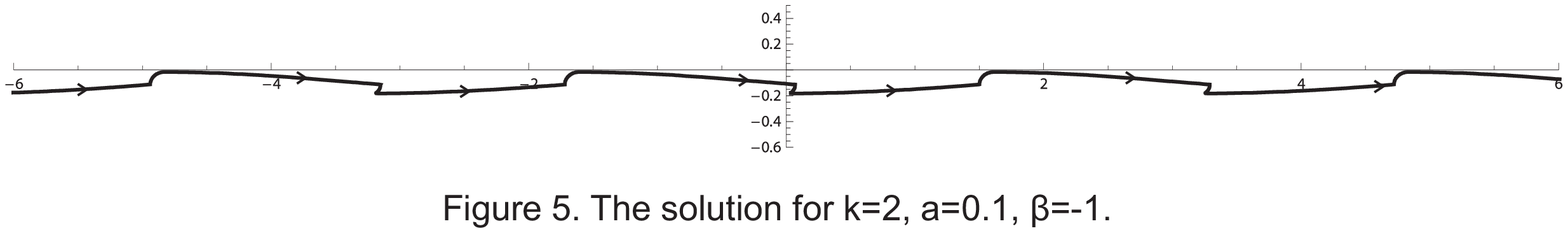}}
\\

We choose now  $k=4$,  $g=9.8$, $a=0.1$ and $\beta=1$. We get by
(\ref{c1}), choosing the square root with sign plus,
  $c=1.56525$,
$A=0.6261$,  but the cubic polynomial under the square root in
(\ref{50'}) has only one real root and we get $Z(t)$ in the form
(\ref{Z2}). The graph of the  curve obtained is depicted in Figure
6. We observe that this solution has vertical asymptotes in the
positive direction.
\\

\hspace{2cm} \scalebox{0.40}{\includegraphics{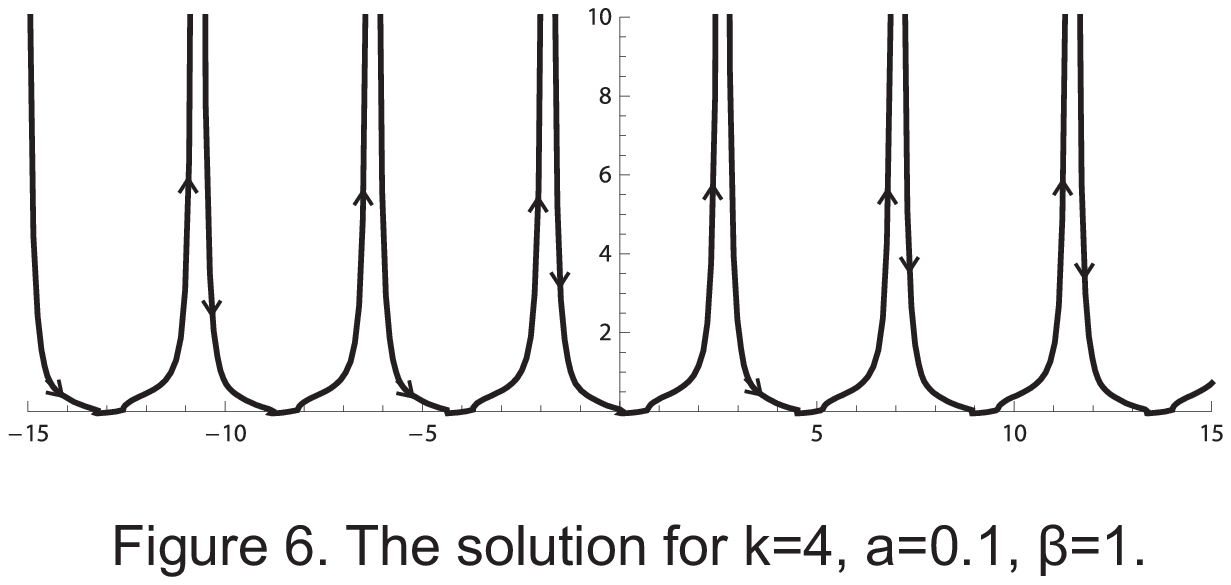}}
\\

\medskip

\medskip

\end{document}